\documentclass[aps,prl,reprint,superscriptaddress,nofootinbib]{revtex4-1}
\usepackage{graphicx}
\usepackage{multirow}
\usepackage{color}

\usepackage{amssymb}
\usepackage{amsmath}
\usepackage{color}
\usepackage{lineno}
\usepackage{enumitem}
\usepackage{comment}
\usepackage[normalem]{ulem}

\newcommand{\br}{{\vec r}}
\newcommand{\bR}{{\vec R}}

\def\nuc#1#2{\relax\ifmmode{}^{#1}{\protect\text{#2}}\else${}^{#1}$#2\fi}

\newcommand{\be}{\begin{eqnarray}}
\newcommand{\ee}{\end{eqnarray}}

\newcommand{\bwt}{\begin{widetext}}
\newcommand{\ewt}{\end{widetext}}

\begin{document}

\title{The puzzle of complete fusion suppression in weakly-bound nuclei: a Trojan Horse effect?}


\author{Jin Lei}
\email[]{jinl@ohio.edu}

\affiliation{Institute of Nuclear and Particle Physics, and Department of Physics and Astronomy, Ohio University, Athens, Ohio 45701, USA}
\affiliation{Departamento de FAMN, Universidad de Sevilla, Apartado 1065, 41080 Sevilla, Spain.}

\author{Antonio M. Moro}
\email[]{moro@us.es}

\affiliation{Departamento de FAMN, Universidad de Sevilla,
Apartado 1065, 41080 Sevilla, Spain.}


\begin{abstract}
Experimental studies of nuclear collisions involving light weakly-bound nuclei show a systematic suppression of the complete fusion cross section by $\sim$30\% with respect to the expectation for tightly bound nuclei, at energies above the Coulomb barrier. Although it is widely accepted that the phenomenon is related to the weak binding of these nuclei, the origin of this suppression is not fully understood. In here,  we present a novel approach that provides the complete fusion for weakly bound nuclei and relates its suppression to the competition between the different mechanisms contributing to the reaction cross section.  The method is applied to the  $^{6,7}$Li+$^{209}$Bi reactions,  where we find that the suppression of complete fusion is mostly caused by the flux associated with non-elastic breakup modes, such as the partial capture of the projectile (incomplete fusion), whereas the elastic breakup mode is found to play a minor role. Finally, we demonstrate that the large  $\alpha$ yields observed in these reactions can be naturally explained as a consequence of a  {\it Trojan Horse} mechanism.


\end{abstract}


\date{\today}%
\maketitle

{\it Introduction}.--
\label{sec:intro}
Fusion between atomic nuclei constitutes a complicated quantum-mechanical dynamical process, whose outcome is critically dictated by a delicate interplay between the coupling of the relative motion of the colliding partners with their internal degrees of freedom. 

Experiments with light weakly-bound stable nuclei (such as $^{6,7}$Li and $^{9}$Be) have shown that the complete fusion (CF)  cross sections   (defined as capture of the complete charge of the projectile) are suppressed by $\sim$20-30\% compared to the case of tightly bound nuclei \cite{Das99,Tri02,Das02,Das04,Muk06,Rat09,LFC15}.  
The effect has been attributed to the breakup of the weakly bound projectile prior to reaching the fusion barrier, with the subsequent reduction of probability of complete  capture. This interpretation is supported by the presence of large $\alpha$ yields as well as target-like residues  which are consistent with the capture of one of the fragment constituents  of the projectile, a process which is usually termed as {\it incomplete fusion} (ICF).   

To account for these observations, some authors have proposed a two-step scenario \cite{Das02,Dia07}  in which the projectile, due to its loosely bound structure,  breaks  into two or more fragments, and then one of them is captured by the target. 
However, dynamical calculations based on a three-dimensional classical dynamical model \cite{Dia07}, which implement this two-step breakup-fusion mechanism, 
can only explain a small fraction of the observed CF suppression for $^{9}$Be \cite{Coo16} and $^{8}$Li \cite{Coo18} reactions. Coupled-channels calculations, including the coupling to low-lying excited states of the projectile and target \cite{Das99,Das02,Kum12,Zha14,Fan15} also fail to describe experimental fusion data. 

Another problem arises in the interpretation of CF of neutron-rich weakly-bound nuclei. In these nuclei, the lowest breakup threshold corresponds to neutron emission. Since CF is operationally defined as capture of the complete charge of the projectile, breakup into one charged fragment and one uncharged one cannot contribute to CF suppression.  Still, for the nucleus $^{8}$Li, whose lowest breakup threshold is $^{7}$Li+$n$ ($S_n=2.03$~MeV),  a large CF suppression of $\sim$30\% has been reported for the $^{8}$Li+$^{208}$Pb  \cite{Agu09} and  $^{8}$Li+$^{209}$Bi \cite{Coo18} reactions.



In this Letter, we propose a novel approach to compute CF cross sections of weakly-bound nuclei. Within a unified fully quantum-mechanical framework, the model is able to explain, simultaneously, the large  $\alpha$-particle yields, the CF cross sections and the connection of their suppression with the binding  energy of the projectile. 

\begin{figure}[tb]
\begin{center}
 {\centering \resizebox*{0.94\columnwidth}{!}{\includegraphics{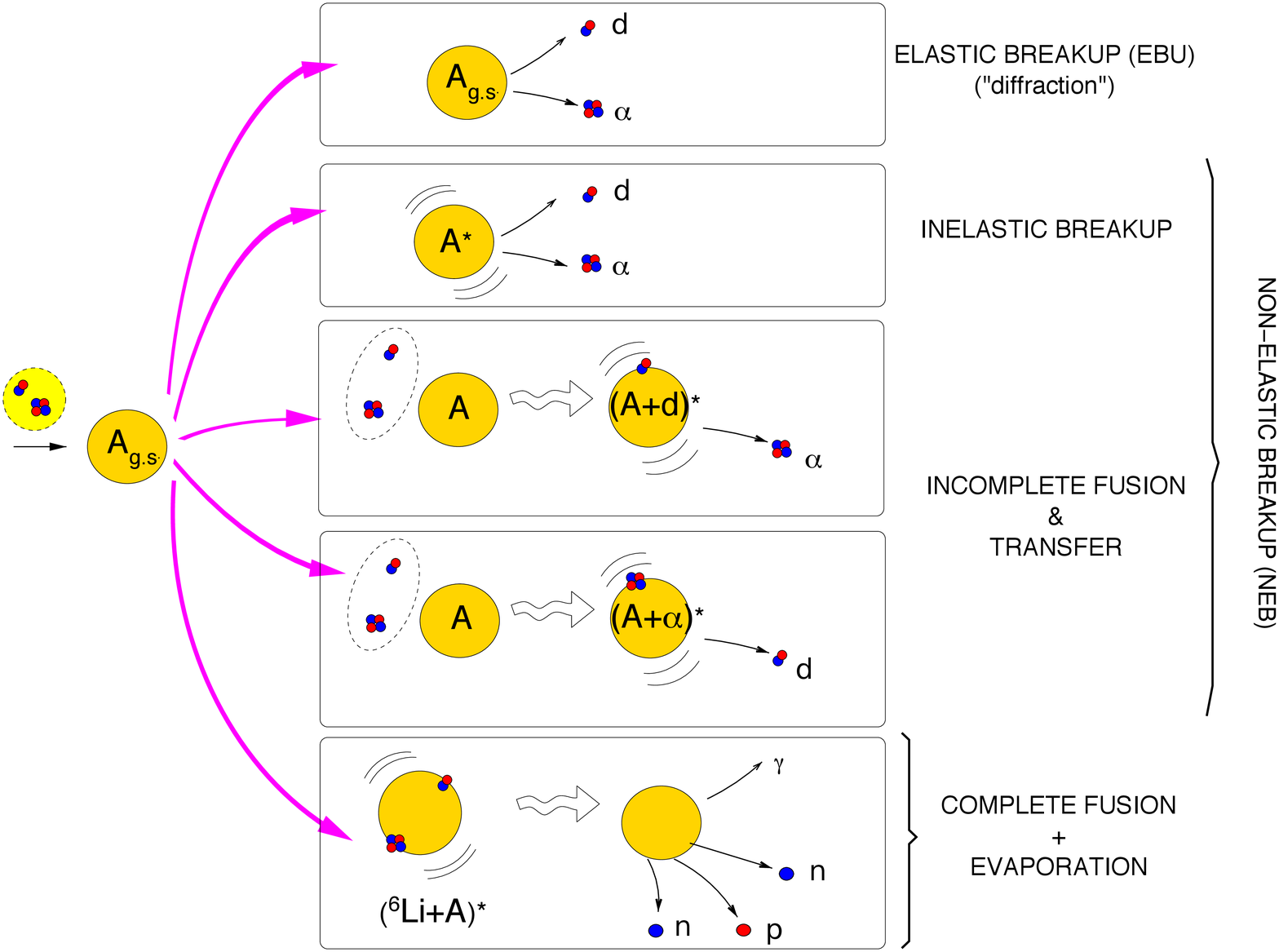}} \par}
\caption{\label{fig:li6dA}Illustration of transfer/breakup modes for a $^6$Li + A reaction. See text for details.}
\end{center}
\end{figure}

{\it  Theoretical framework}.--
 \label{sec:theo}
We consider a collision of a weakly-bound two-body projectile (denoted $a=b+x$) with a target nucleus $A$. 
We are mainly concerned here with the process in which the projectile as a whole fuses with the target nucleus, that is, {\it complete fusion} (CF). A realistic evaluation of the CF cross section must take into account the effect of other  channels, such as projectile and/or target excitation, transfer and breakup. 
The explicit inclusion of all these channels in actual calculations is however not possible due to the overwhelming number of processes involved. To overcome this difficulty, the model proposed here  takes advantage of the fact that light, weakly-bound nuclei have a marked cluster structure which suggests a natural decomposition of non-elastic channels in terms of the processes undergone by each of the clusters. Furthermore, the sum of the CF plus the other non-elastic channels is a well-constrained quantity since it is given by the reaction cross section ($\sigma_R$). Consequently, for a two-body projectile  we may write the following approximate decomposition  
 \begin{equation}
 \label{eq:decomp}
 \sigma_R \approx    \sigma_\mathrm{CF} + \sigma_\mathrm{inel} + \sigma_\mathrm{EBU} +
 \sigma^{(b)}_\mathrm{NEB} + \sigma^{(x)}_\mathrm{NEB} .
 \end{equation}
In this expression, $\sigma_\mathrm{inel}$ corresponds to the excitation of the projectile and/or target without dissociation (i.e., inelastic scattering). The term $\sigma_\mathrm{EBU}$  corresponds to {\it elastic breakup}, defined as the dissociative processes in which {\it both} fragments interact elastically with the target nucleus and hence the three outgoing fragments are emitted in their ground state (i.e., $a + A \rightarrow b + x + A_\mathrm{gs}$).  Finally, $\sigma^{(b)}_\mathrm{NEB}$ and $\sigma^{(x)}_\mathrm{NEB}$ account for the so-called {\it non-elastic breakup} (NEB) processes, in which one of the two fragments  interacts non-elastically with the target nucleus. This includes the ICF  described in the introduction but also other processes, such as the projectile dissociation accompanied by target excitation ($a + A \rightarrow b + x + A^*$) or the exchange of nucleons between one the projectile fragments and the target.  
The outlined processes are schematically depicted  in Fig.~\ref{fig:li6dA} using as example a $^{6}$Li+A reaction (modeled as  $\alpha+d + A$).

The central idea of the present method is that the quantities $\sigma_R$, $\sigma_\mathrm{inel}$, $\sigma_\mathrm{EBU}$  and $\sigma^{(b,x)}_\mathrm{NEB}$ can be reliably calculated with existing reaction formalisms so that the CF section can be inferred from Eq.~(\ref{eq:decomp}). The pure inelastic scattering cross sections ($\sigma_\mathrm{inel}$) are standardly computed by means of coupled-channels calculations including low-lying collective excitations of the projectile and target.  The EBU part  can be accurately calculated using the continuum-discretized coupled-channels (CDCC) method \cite{Aus87}, which treats the breakup as an excitation to the continuum states. Evaluation of the non-elastic breakup modes is much more challenging because of the large number of processes involved. Here, we propose to use the spectator/participant inclusive breakup model of Ichimura, Austern and Vincent (IAV) \cite{Aus81,Aus87,IAV85},  in which the explicit sum over final states arising from the interaction of the participant particle with the target is avoided by using the Feshbach projection formalism, giving rise to a closed-form formula for the the double differential cross section for  NEB  with respect to the angle and energy of the spectator fragment. For example, if $x$ is the participant particle,
\begin{equation}
\label{eq:iav_3b}
\left . \frac{d^2\sigma}{dE_b d\Omega_b} \right |_\mathrm{NEB} = -\frac{2}{\hbar v_{a}} \rho_b(E_b)  \langle \varphi_x (\vec{k}_b) | \mathrm{Im}[U_{xA}] | \varphi_x (\vec{k}_b) \rangle   ,
\end{equation}
where $\rho_b(E_b)$ is the density of states of the particle $b$,
$v_a$ is the velocity of the incoming particle, $U_{xA}$ is the optical potential describing $x+A$ elastic scattering, and  $\varphi_x(\vec{k}_b,\br_{xA})$ is a projected wave function  describing the evolution of the $x$ particle after dissociating from the projectile, when the core is scattered with momentum $\vec{k}_b$. This function is obtained 
from the equation
$\varphi_x(\vec{k}_b,\br_{xA}) =\int G^\text{opt}_x (\br_{xA},\br'_{xA}) \langle \br'_{xA}\chi_b^{(-)}| V_\text{post}|\Psi^{3b} \rangle d\br'_{xA}$, where $G^\text{opt}$ is the optical model Green's function with potential $U_{xA}$, 
%
$\chi_b^{(-)}(\vec{k}_b,\br_{bB})$ is the distorted-wave describing the scattering of the outgoing $b$ fragment  with respect to the $B\equiv x+A$ system (obtained with some optical potential $U_{bB}$),  $V_\mathrm{post} \equiv V_{bx}+U_{bA}-U_{bB}$ is the post-form transition operator and $\Psi^{3b}$ the three-body scattering wave function. Further details can be found in Ref.~\cite{Jin15}. 
Following our previous works \cite{Jin15,Jin15b,Jin17},
we approximate $\Psi^{3b}$ by its  DWBA form: $\Psi^{3b}(\bR,\br) \approx \chi^{(+)}_{a}(\bR) \phi_{a}(\br)$, where $\chi^{(+)}_{a}(\bR)$ is a distorted wave describing $a+A$ elastic scattering, obtained with some optical potential, and  $\phi_{a}(\br)$ is the projectile ground state wave function. 
Notice that the  expectation value of the imaginary part of the $U_{xA}$ potential in Eq.~(\ref{eq:iav_3b}) accounts for all possible non-elastic processes which may take place in $x-A$ scattering (that is, NEB), no matter how  diverse or complicated they are. 
Recent applications of this DWBA version of the IAV model to deuteron \cite{Pot15,Carlson2016,Jin15}, $^6$Li \cite{Jin15,Jin17}, and  $^{7}$Li  \cite{Jin18b} induced reactions have shown a very good agreement with existing data.  We note that, although a  decomposition similar to Eq.~(\ref{eq:decomp}) has been employed by other authors \cite{Par16}, a proper computation of the NEB contributions, using a well founded theory, is a key and novel aspect of the present approach.

Finally, the reaction cross section ($\sigma_R$) can be extracted using the elastic S-matrix from the CDCC calculation or from an optical model fit of the elastic data, if available.  

 

\begin{figure}[tb]
\begin{center}
 {\centering \resizebox*{0.85\columnwidth}{!}{\includegraphics{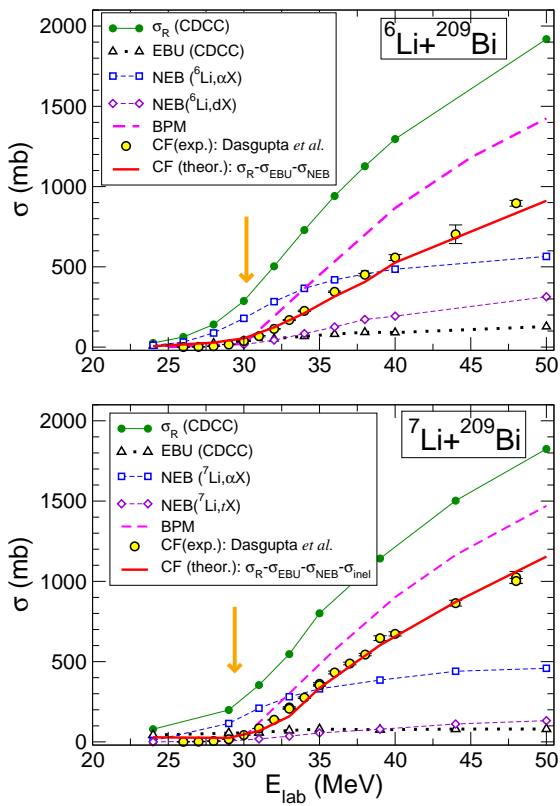}} \par}
\caption{\label{fig:li6bi_cf} Complete fusion cross section for the $^{6,7}$Li+$^{209}$Bi reactions as a function of the incident energy.  Experimental data are from Ref.~\cite{Das04}. 
The arrow indicates the nominal position of the Coulomb barrier.}
\end{center}
\end{figure}

{\it Application to the $^{6,7}$Li+$^{209}$Bi reactions}.--
We apply now the proposed methodology  to the reactions $^{6,7}$Li+$^{209}$Bi. CF cross sections for these reactions have been measured by Dasgupta {\it et al.}~\cite{Das02,Das04}, at energies below and above the Coulomb barrier ($V_b \approx 30$~MeV), and their results are shown in Fig.~\ref{fig:li6bi_cf} (yellow circles), with the top and bottom panels corresponding to the $^6$Li and $^7$Li cases, respectively. 
 CF suppression is usually measured with respect to the single-barrier penetration model (BPM), which accounts for the quantum tunneling probability through the effective Coulomb plus centrifugal barrier but ignoring the effect of other channels. These BPM calculations (quoted from Ref.~\cite{Das02}), are shown by magenta dashed lines. The effect of CF suppression is clearly apparent, amounting to $\sim$30\% and $\sim$25\% for the  $^6$Li and $^7$Li cases, respectively.

To evaluate the CF cross section in presence of the other non-elastic channels, we make use of Eq.~(\ref{eq:decomp}). The projectile inelastic scattering  and EBU cross sections are obtained from CDCC calculations, using a two-body  model ($\alpha+x$, with $x=d$ or $x=t$) for $^{6,7}$Li. For the $^6$Li+$^{209}$Bi case, these calculations follow closely those performed in Ref.~\cite{Jin15}, so we refer  to this work for further details. For the $^7$Li+$^{209}$Bi reaction, we employ the $\alpha+t$ model parameters from Ref.~\cite{Buck88}  and the $t$-target and $\alpha$-target potentials from 
Refs.~\cite{BG71} and \cite{Barnett74}, respectively. 
Following our previous works \cite{Jin15,Jin17}, the   $d$-target and $t$-target potentials are renormalized to better reproduce the corresponding $^{6,7}$Li+$^{209}$Bi  elastic cross sections. 
 Target excitations were not considered, since they have been shown to  have a negligible effect on fusion at the above-barrier energies considered here. 

The NEB cross sections are computed with the DWBA version of the IAV model described above. Within our assumed two-body model of $^{6,7}$Li, there are two distinct contributions, namely, one in which $x$ interacts non-elastically with the target (with $\alpha$ acting as a spectator) and another in which  $\alpha$ interacts non-elastically. The same potentials are used in both calculations, and just the roles of participant and spectator are exchanged in Eq.~(\ref{eq:iav_3b}).
These $\alpha$ and $x$ yields are displayed, respectively, by squares and diamonds in Fig.~\ref{fig:li6bi_cf}. In \cite{Jin15}, we showed that these calculations reproduce very well the inclusive $\alpha$ distributions measured in Ref.~\cite{Santra12} for the $^{6}$Li+$^{209}$Bi reaction.

Finally, the reaction cross sections were evaluated from the elastic S-matrices obtained from the CDCC calculations. These reaction cross sections were found to be very close to those computed with the optical model fit of the elastic cross section from Refs.~\cite{Santra12,Martel95}. 

It is seen in Fig.~\ref{fig:li6bi_cf} that the calculated  CF cross sections  (red solid lines), deduced from Eq.~(\ref{eq:decomp}), are remarkably close to the data. The separate role of each of the competing channels can be also deduced  from this figure.  
The EBU mechanism ($\alpha+d$ and $\alpha+t$ production) plays a minor role, representing a small fraction of the reaction cross section at the incident energies relevant for this work. Instead, the dominant breakup mechanism in both reactions is the $\alpha$ production due to the ($^{6,7}$Li,$\alpha$$X$) NEB. This explains the large $\alpha$ yields observed experimentally in these reactions. This is in fact a rather general feature found independently of the target nucleus \cite{Jin17}. 

The deuteron-production ($^{6}$Li,$d$$X$) and triton-production ($^{7}$Li,$t$$X$) NEB channels are much smaller than the $\alpha$-production ones. This can be understood as a combination of two effects: (i) the lower Coulomb barrier energy {\it felt} by the $d$ and $t$ particles as compared to the $\alpha$ particle and (ii) the smaller reaction cross section for the $\alpha$ particles, owning to its tightly-bound, compact structure.


 The fact that the EBU mechanism barely affects the CF cross section  explains why classical  \cite{Coo16} and quantum-mechanical calculations \cite{Elm15}, which  
 consider the fusion suppression due to the population of these elastic breakup channels, can only account for a small fraction of this suppression. 

\begin{figure}[tb]
\begin{center}
 {\centering \resizebox*{0.94\columnwidth}{!}{\includegraphics{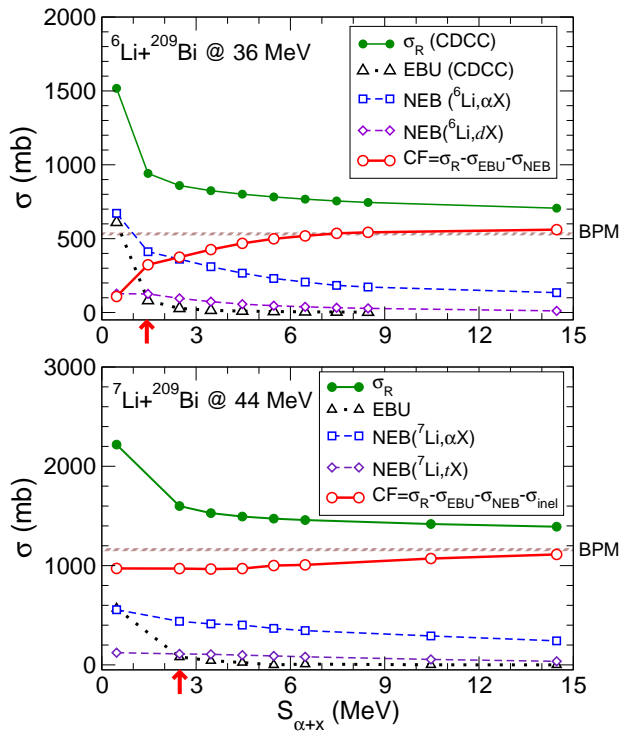}} \par}
\caption{\label{fig:sup} Dependence of EBU, NEB and CF cross sections on the separation energy for $^{6,7}$Li, $S_{\alpha+x}$ ($x=\alpha,t$) for  $^{209}$Bi($^{6}$Li,\,$\alpha$$ X$) (upper panel) and $^{209}$Bi($^{7}$Li,\,$\alpha$X) (lower panel) reactions. The vertical arrow indicates the physical separation energy. The BPM limit is shown with an horizontal gray line.}
\end{center}
\end{figure}

Although direct breakup plays a minor role in CF suppression, the degree of  suppression has been shown to be  closely correlated with the separation energy of the projectile into its cluster constituents \cite{Gas09}. 
To investigate this connection within the present framework, 
we have repeated the calculations varying artificially the separation energy of the  $^6$Li and $^7$Li nuclei for selected  incident energies. The results are shown in Fig.~\ref{fig:sup} for  $^6$Li + $^{209}$Bi at 36~MeV (1.2$V_b$) and $^7$Li + $^{209}$Bi at 44~MeV (1.5$V_b$). For each case, the BPM limit is indicated by a horizontal line. It is seen that, as the separation energy is increased with respect to its physical value, the reaction cross section decreases monotonically, indicating an overall reduction of non-elastic channels, as expected. The EBU contribution falls very fast, becoming negligible for separation energies of $\sim$$3$-$4$~MeV. The NEB contributions decrease also with the separation energy, but at a much lower rate, particularly for the $x$-fragment absorption. Interestingly, for large separation energies the difference $\sigma_R  - \sigma_\mathrm{EBU} - \sigma_\mathrm{NEB}-\sigma_\mathrm{inel}$, that in our model is identified with $\sigma_\text{CF}$, tends to the BPM values for both the  $^6$Li and $^7$Li cases. Thus, in the limit of strong binding, our model predicts no suppression, as expected. This reinforces our interpretation that the CF suppression arises from the flux associated with the transfer/breakup modes due to the weakly-bound structure of the projectile. 

\begin{figure}[tb]
\begin{center}
 {\centering \resizebox*{0.8\columnwidth}{!}{\includegraphics{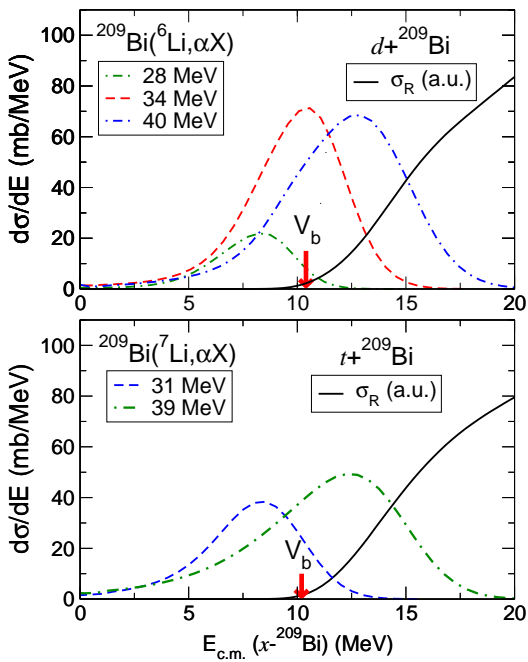}} \par}
\caption{\label{fig:thm} Illustration of the Trojan Horse mechanism for the $^{209}$Bi($^{6}$Li,\,$\alpha$$X$) (upper panel) and $^{209}$Bi($^{7}$Li,\,$\alpha$$X$) (lower panel) reactions. The broken lines are the three-body cross sections ($^{6,7}$Li,\,$\alpha$$X$) at the  incident energies indicated by the labels. The black solid lines are the two-body reaction cross sections (in arbitrary units). The nominal position of the barrier for the $d$+$^{209}$Bi and $t$+$^{209}$Bi reactions is indicated by the arrow. }
\end{center}
\end{figure}

The calculations just   presented rule out the direct breakup ($^6$Li$\to \alpha+d$ and $^7$Li$\to \alpha+t$) and  point toward the $\alpha$-production NEB mechanisms as the main responsible mechanism for the CF suppression in $^{6,7}$Li-induced reactions. As noted earlier, these channels are associated, respectively, with deuteron and triton reactions with the target nucleus. This includes particle transfer, target excitation and ICF. This may seem unexpected 
 if one notes that the average deuteron and triton kinetic energies in the incident $^{6}$Li and $^{7}$Li projectiles are of the order, or even smaller, than their respective Coulomb barrier energies for the $d$+$^{209}${Bi} and $t$+$^{209}${Bi} systems ($\sim$10-11~MeV). For such low  incident energies, the free $d$+$^{209}${Bi} and $t$+$^{209}${Bi}  reaction cross sections are very small, in spite of which, the three-body  $^{209}$Bi($^{6,7}$Li,\,$\alpha$$X$) cross sections are remarkably large. This phenomenon is not new and was first pointed out by Baur \cite{Bau86}, who explained it invoking a ``Trojan Horse mechanism''. The idea is that, for a three-body reaction of the form $a+A$, with $a=b + x$, a particular channel of the form $a+A \to b+ c +C$ will be enhanced  
with respect to the free, two-body reaction $x+A \to c + C$ due to the fact that the $a+A$ system is above its Coulomb barrier. Loosely speaking, the $x$ particle is brought inside its Coulomb barrier by the heavier particle $a$.
 The method has become a standard tool in nuclear astrophysics as an indirect way of obtaining information of low-energy charged-particle induced reactions by means of three-body reactions (see e.g.~\cite{Spi11}) and its formal aspects can be found elsewhere \cite{Typ03}. We illustrate here the phenomenon for the two reactions under study. For that, in Fig.~\ref{fig:thm} we compare the reaction cross sections for the {\it two-body} reactions $d$+$^{209}${Bi} and $t$+$^{209}${Bi}, as a function of the center-of-mass energy for each system, with the {\it three-body} cross sections  $^{209}$Bi($^{6}$Li,\,$\alpha$$X$) (top) and $^{209}$Bi($^{7}$Li,\,$\alpha$$X$) (bottom) for several $^{6,7}$Li incident energies. The vertical arrow in each panel denotes the position of the Coulomb barrier for the $d/t$+$^{209}${Bi} system. As expected, the reaction cross section for the two-body reactions drops very quickly as the energy approaches the Coulomb barrier. By contrast, the three-body cross sections remain very large, even at energies well below  their nominal barrier. These results provide a natural explanation of the large  $\alpha$  yields observed experimentally and confirmed by the IAV model.

 The picture that emerges from these calculations is the following. The weakly bound projectile $a$ overcomes the $a+A$ Coulomb barrier, bringing also the $x$ fragment inside its Coulomb barrier via the just described Trojan Horse mechanism. This triggers the non-elastic processes between $x$ and $A$ which give rise to the large variety of emerging fragments observed experimentally and, in turn, to the suppression of CF. The present results add numerical support to the suggestion put forward by Cook {\it et al.}~\cite{Coo18}, who conjectured that it is clustering and weak-binding, but not breakup in the usual sense, that is responsible for 
the CF suppression.

{\it Summary and conclusions}.--
 \label{sec:sum}
In summary, we have proposed a new method to compute CF cross sections in collisions of light, weakly-bound nuclei. The method links these cross sections with the reaction and the transfer/breakup cross sections. These quantities can be reliably evaluated with state-of-the-art reaction frameworks, namely, the CDCC method for the EBU part, and the  inclusive breakup model of IAV for the NEB. Application to the $^{6,7}$Li + $^{209}$Bi reactions, shows an excellent agreement with the CF data for these systems, and shows that the CF suppression originates from the flux associated with non-elastic breakup modes, most notably the $\alpha$ production channels. The large yields observed for these channels can be naturally explained as due to a {\it Trojan Horse}  mechanism. 
Contrary to the assumption made in some works, we find that the direct breakup channels ($^{6}\text{Li} \to \alpha+d$  and $^{7}\text{Li}\to \alpha+t$), which can be identified with our EBU contribution, play a very small role for these systems. 


 Although the calculations presented here have been restricted to the $^{6,7}$Li projectiles, we expect the conclusions to be valid for other weakly bound nuclei for which CF suppression have been also reported, such as $^{9}$Be or $^{8}$Li. 
 An interesting question that arises is how the relative importance of the different competing mechanisms evolve as the separation energy of the projectile decreases, such as in the extreme cases of the halo nuclei $^{11}$Li, $^{6}$Li or $^{11}$Be.  We note that, although the methodology proposed here is in principle applicable  to these more exotic systems, its application may require (i) going  beyond the DWBA approximation adopted here for the NEB cross sections and (ii), in the case of  $^{11}$Li and $^{6}$He, a description of the projectile in terms of a three-body cluster model. 
 

\bigskip
\begin{acknowledgments}
We are grateful to Daniel Phillips, Mahananda Dasgupta and Ed Simpson for a critical reading of the manuscript and many  insightful comments on this work.  This work has been partially supported by the National Science Foundation
under contract No.\ NSF-PHY-1520972 with Ohio University,
by the Spanish Ministerio de Ciencia, Innovaci\'on y Universidades and FEDER funds (projects FIS2014-53448-C2-1-P and FIS2017-88410-P)   and by the European Union's Horizon 2020 research and innovation program under grant agreement No.\ 654002.
 This research used resources of the National Energy Research Scientific Computing Center (NERSC), a U.S. Department of Energy Office of Science User Facility operated under Contract No. DE-AC02-05CH11231.
\end{acknowledgments}

\bibliography{inclusive_prc.bib}
\end{document}